\documentclass[%
aip,
amsmath,amssymb,
reprint,%
]{revtex4-1}

\usepackage{graphicx}
\usepackage{dcolumn}
\usepackage{bm}

\usepackage[utf8]{inputenc}
\usepackage[T1]{fontenc}
\usepackage{mathptmx}

\begin{document}

\preprint{AIP/123-QED}

\title{Generation of pure-state single photons with high heralding efficiency by using a three-stage nonlinear interferometer}

\author{Jiamin Li}
 \affiliation{College of Precision Instrument and Opto-Electronics Engineering, Key Laboratory of Opto-Electronics Information Technology, Ministry of Education, Tianjin University, Tianjin 300072, P. R. China}

\author{Jie Su}
 \affiliation{College of Precision Instrument and Opto-Electronics Engineering, Key Laboratory of Opto-Electronics Information Technology, Ministry of Education, Tianjin University, Tianjin 300072, P. R. China}

\author{Liang Cui}%
 \email{lcui@tju.edu.cn}
\affiliation{College of Precision Instrument and Opto-Electronics Engineering, Key Laboratory of Opto-Electronics Information Technology, Ministry of Education, Tianjin University, Tianjin 300072, P. R. China}

\author{Tianqi Xie}%
\affiliation{College of Precision Instrument and Opto-Electronics Engineering, Key Laboratory of Opto-Electronics Information Technology, Ministry of Education, Tianjin University, Tianjin 300072, P. R. China}

\author{Z.Y. Ou}
\affiliation{College of Precision Instrument and Opto-Electronics Engineering, Key Laboratory of Opto-Electronics Information Technology, Ministry of Education, Tianjin University, Tianjin 300072, P. R. China}%
\affiliation{Department of Physics, Indiana University-Purdue University Indianapolis, Indianapolis, IN 46202,USA}

\author{Xiaoying Li}
 \email{xiaoyingli@tju.edu.cn}
\affiliation{College of Precision Instrument and Opto-Electronics Engineering, Key Laboratory of Opto-Electronics Information Technology, Ministry of Education, Tianjin University, Tianjin 300072, P. R. China}%

\date{\today}

\begin{abstract}
We experimentally study a fiber-based three-stage nonlinear interferometer and demonstrate its application in generating heralded single photons with high efficiency and purity by spectral engineering. We obtain a heralding efficiency of $90\%$ at a brightness of 0.039 photons/pulse. The purity of the source is checked by two-photon Hong-Ou-Mandel interference with a visibility of $95\pm 6\% $ (after correcting Raman scattering and multi-pair events). Our investigation indicates that the heralded source of single photons produced by the three-stage nonlinear interferometer has the advantages of high purity, high heralding efficiency, high brightness, and flexibility in wavelength and bandwidth selection.
\end{abstract}

\maketitle


Single photons are a fundamental resource for quantum optics and quantum information processing (QIP) \cite{kumar2004QIP}. Single-photon state can be obtained by heralding on the detection of one of the photon pairs generated from spontaneous parametric processes\cite{mandel-PRL86}.
Even with the advancement in producing good quality “on-demand” single-photon sources \cite{Santori02,Ding16}, heralded single-photon sources are proven to be effective, economical, and most importantly of high quality.

The indistinguishability or modal purity is the most desired feature of a single-photon source  and can be measured by Hong-Ou-Mandel (HOM) interference between two independent photons \cite{ou99}. This interference phenomenon plays a key role in QIP, such as quantum teleportation \cite{Bou97,duan2001} and linear optical quantum computing \cite{Knill01}. When the multi-photon events of signal and idler photon pairs are negligible, the visibility $V$ of HOM interference between independent sources $V$ and the indistinguishability of a heralded single-photon source characterized by the mode number $M$ is related via $V=\frac{1}{M}$. For single photons in a well defined single-mode, i.e., $M=1$, the ideal visibility with $V=1$ is achievable.
Another important criterion for characterizing the performance of single photons is the photon statistics measured by the second-order correlation function $g^{(2)}$. An ideal single-photon source corresponds to $g^{(2)}=0$. For the source of heralded single photons, we have $g^{(2)}=\frac{2R_c}{h_sh_i}(1+\frac{1}{M})$,\cite{su-arxiv} where $R_c$ denotes the rate or brightness of single photons, $h_{s(i)}$ is referred to as the heralding efficiency or collection efficiency, which describes the probability of a photon emerging at signal (idler) band upon the detection event in idler (signal) band. Therefore, to develop the source of heralded single photons, both modal purity $M \rightarrow 1$ and high collection efficiency $h_sh_i \rightarrow 1$ are desirable.

For the pulse-pumped spontaneous parametric process, the precise timing provided by the ultra-short pump pulses brings convenience for synchronizing independent sources. However, the broadband nature of the pump field and strict phase matching condition in highly dispersive nonlinear medium lead to complicated spectral correlation in frequency domain, which will degrade the purity of single photons \cite{ou99}. To avoid this degradation, tremendous efforts were spent through the years. A straightforward approach is to apply passive narrow band filters to enforce single mode operation \cite{Ou97}, but at the cost of the reduction in brightness and collection efficiency $h_sh_i\ll1$. Another approach is to specially engineer the nonlinear media with just the required dispersion to achieve single-mode operation without using filter. In this case, $M\approx1$ and $h_sh_i\rightarrow 1$ are simultaneously attainable \cite{Grice-PRA-2001,Mosley08,cui2012NJP,Garay07}. However, because the nonlinear interaction and linear dispersion in nonlinear media are often mixed in parametric processes, limited successes have been achieved so far only at some specific wavelengths with sophisticated design \cite{evans10,Branczyk11OE}.

Recently, our group adopted a new approach of engineering the quantum states with a nonlinear interferometer (NLI) \cite{su-arxiv,Su19}. In a proof-of-principle experiment, using a two-stage NLI formed by two identical nonlinear fibers with a linear medium of single mode fiber (SMF) in between, we demonstrated that the spectra of photon pairs was engineered to nearly factorable states without sacrificing collection efficiency. Exploiting the quantum interference of two pulse-pumped spontaneous four wave
mixing (FWM) processes in nonlinear fibers, the spectra of photon pairs can be flexibly reshaped by dispersion engineering in the linear medium of SMF without affecting the phase matching condition of FWM \cite{su-arxiv}.

Furthermore, we showed theoretically that even better properties of the photon source can be achieved by implementing a multi-stage NLI (stage number $N>2$) by a finer control of the phase shift of quantum interference \cite{Su19,su-arxiv}. Although such an idea was first proposed and analyzed by U’Ren {\it et al.}\cite{uRen2005}, the proposed nano-structure of nonlinear crystal is not easy to implement experimentally. In this paper, we experimentally realize a three-stage NLI with optical fibers, which make it relatively easy to extend to higher stage number, and perform a two-photon HOM interference with the photon sources so produced to verify the high quality of the source.

Figure 1(a) shows the conceptual sketch of our three-stage NLI, consisting of three identical dispersion shifted fibers (DSF) with two identical standard SMFs placed in between the DSFs. The DSFs with length $L$ serve as nonlinear media of spontaneous FWM, while the SMFs with length $L_{DM}$
function as the linear dispersive media with which we can engineer the spectrum of the generated photons.
When the pump launched into the NLI is a Gaussian-shaped pulse trains with central frequency and bandwidth of $\omega_{p 0}$ and $\sigma_{p}$, respectively, the joint spectra function (JSF) of the signal and idler photon pairs ($\omega_s,\omega_i$) emerging at the output of NLI is expressed as \cite{su-arxiv}:
\begin{equation}\label{NLI_multistage}
\begin{split}
&F_{N L I}\left(\omega_{s}, \omega_{i}\right)\\
&=\exp \big[-\left(\omega_{s}+\omega_{i}-2 \omega_{p 0}\right)^{2}/4 \sigma_{p}^{2}\big] \operatorname{sinc}\left(\Delta k L/2\right) H(\theta),
\end{split}
\end{equation}
with $\Delta k =\frac{k^{(2)}}{4} (\omega_s-\omega_i)^2
-2 \gamma P_p$
denoting the phase matching function in each DSF, where $H(\theta ) = e^{j(N - 1)\theta }\sin N\theta /\sin \theta$ results from interference for N-stage NLI ($N=3$ for our case), $k^{(2)} = \frac{\lambda^2_{p0}}{2\pi c} D_{slope}(\lambda_{p0}-\lambda_{0})$ is the second order dispersion coefficient of DSF, and $D_{slope}$ is dispersion slope of DSF at zero dispersion wavelength (ZDW) $\lambda_0$. $\gamma$ is the nonlinear coefficient of DSF, $P_p$ is the peak power of pump, 
$\theta  = \frac{1}{2}(\Delta kL + \Delta {\phi _{DM}})$ is the overall phase difference between pump and photon pairs of adjacent DSFs, with $\Delta \phi_{DM}$  being the part contributed by each linear medium.

The key in our design is that the phase difference $\Delta \phi_{DM}$ induced by SMF is much greater than the phase mismatching term in DSF or $\Delta k L/2 \ll 1$, so that the approximation
\begin{equation}
\theta \approx	 {\phi_{DM}}/{2} =  {\lambda_{p 0}^{2} D_{S M F} L_{D M}}(\omega_{s}-\omega_{i})^{2}/{16 \pi c}
\end{equation}
holds, where $D_{SMF}=17~ {\rm ps/(nm\cdot km)}$ is the dispersion parameter of SMF at pump central wavelength $\lambda_{p 0}$. In this case, the interference term $H(\theta)$ is dominated by the linear media.

\begin{figure}[htb]
	\centering
	\includegraphics[width=0.48 \textwidth]{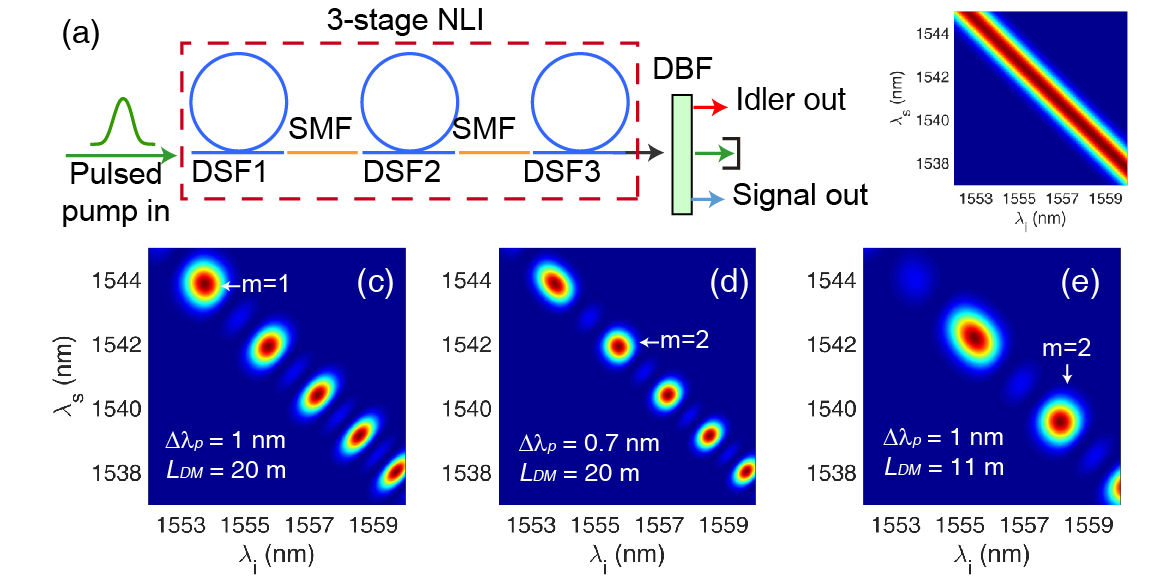}
	\caption{(a) Conceptual sketch of a three-stage nonlinear interferometer (NLI) formed by dispersion-shifted fibers (DSF) and standard single-mode fibers (SMF). (b) The joint spectral intensity (JSI) of photon pairs directly out of DSF1 without interference. The three plots in the bottom are the JSI of photon pairs when the pump bandwidth and SMF length of the NLI are (c) $\Delta \lambda_p$ = 1 nm and $L_{DM}$ = 20 m; (d) $\Delta \lambda_p =0.7$ nm and $L_{DM}=20$ m; (e) $\Delta \lambda_p =1$nm and $L_{DM}=11$ m, respectively.
}
	\label{fig_NLI_photon pair}
\end{figure}


To see the effect of interference, we simulate the spectral property of photon pairs by substituting the key parameters of our NLI into Eqs. (1) and (2). In the calculation, we set the parameters of DSF as: $L=150$ m, $\lambda_0$ =1548.5 nm, $D_{slope}=0.075$ $\rm{ps/(km\cdot nm^2)}$ and $\gamma =2$ $\rm{W\cdot km ^{-1}}$. The central wavelength and peak power of pump are $\lambda_{p0}$ =1548.8 nm and $ P_p = 0.35 $ W.
The joint spectrum intensity (JSI) of photon pairs in Figs. \ref{fig_NLI_photon pair}(c) - (e) is obtained when the pump bandwidth and SMF length of the NLI are (c) $\Delta \lambda_p $ = 1 nm and $L_{DM}$ = 20 m; (d) $\Delta \lambda_p =0.7$ nm and $L_{DM}$ = 20 m; (e) $\Delta \lambda_p =1$ nm and $L_{DM}=11$ m, respectively. Compared to Fig.\ref{fig_NLI_photon pair}(b) without interference, the JSI is broken into islands with various shapes depending on pump bandwidth $\Delta \lambda_p$ and the length of the linear dispersive medium $L_{DM}$.
We can then select the roundest one ($m=1$ in (c) and $m=2$ in (d,e)) for a factorized JSF with a filter of proper center wavelength and bandwidth.  The results indicate that by changing the pump bandwidth and the length of SMF, the wavelength and bandwidth of pure state single photon with $h_sh_i\rightarrow 1$ can be flexibly adjusted within the gain bandwidth of FWM in DSF.

Good separation between the islands is desirable in order to filter out each island \cite{Su19,su-arxiv}. However, the relative separation in Fig. \ref{fig_NLI_photon pair}(c,d,e) do not change with $\Delta \lambda_p$ and $L_{DM}$ but will increase with the stage number $N$, as shown in Ref.[\onlinecite{Su19,su-arxiv}]. Here we see well-separated islands even with $N=3$, which is already better than the case of $N=2$\cite{Su19}.
Note that U’Ren {\it et al.}\cite{uRen2005} had previously proposed the idea of generating pure state single photon by using a sequential array of $\chi^{(2)}$-crystal gapped in between by linear dispersive media. However, since the amount of dispersion induced by each linear spacer and $\chi^{(2)}$ crystal is comparable, the approximation in Eq. (2) is not valid so that it requires a large number of $\chi^{(2)}$-crystals (large $N$ in Eq. (1)) to separate different islands in JSF similar to Fig. \ref{fig_NLI_photon pair}(c,d,e). Because of the sophisticated structure, the proposal in Ref. [\onlinecite{uRen2005}] has not been experimentally realized yet.

\begin{figure*}[htbp]
	\centering
	\includegraphics[width=0.92\textwidth]{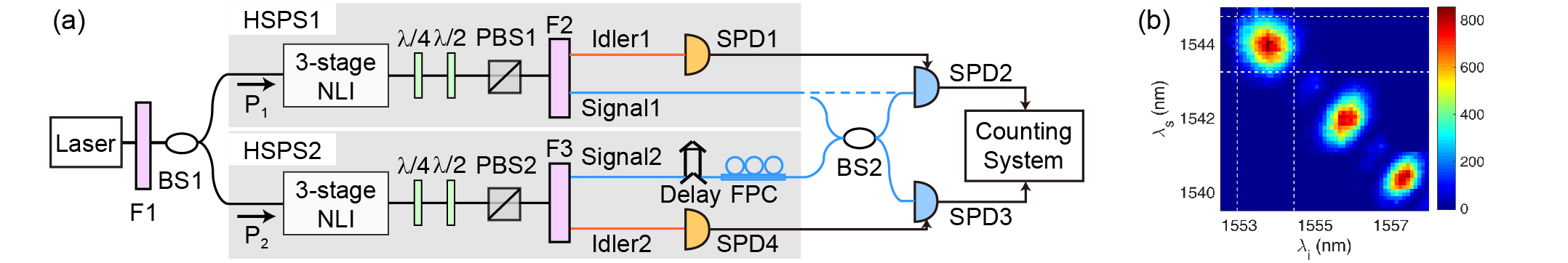}
	\caption{(a) Experimental setup of two-photon HOM interference by using two heralded single photon sources (HSPS1 and HSPS2). The shaded areas are HSPS1 and HSPS2, whose detail is given in Fig. 1(a). P1-P2, pulsed pump; NLI, nonlinear interferometer; F, filter; DBF, dual-band filter; FPC, fiber polarization controller; BS, 50/50 beam splitter; SPD, single photon detector.
	(b) Contour map of true coincidence of photon pairs as function of $\lambda_s$ and $\lambda_i$, reflecting the joint spectral intensity of photon pairs.
	}
	\label{fig_setup}
\end{figure*}

The schematic of the experimental setup is shown in Fig. 2(a) where the gray shaded frames are the heralded single photon sources (HSPS1, HSPS2). The details of each HSPS is given in Fig. 1(a). The parameters of each HSPS are the same as those in plotting Fig. 1(c). The transmission loss of NLI is mostly due to imperfect fiber splicing and is about 17$\%$. The NLI is submerged in liquid nitrogen to suppress Raman scattering  \cite{XYLi04}. The pump P$_1$ or P$_2$ is obtained by passing the output of a femoto-second fiber laser through a bandpass filter (F1) and 50/50 beam splitter (BS1). The repetition rate of laser is about 36.8 MHz. The pulse duration of the pump with FWHM of 1 nm is about 4 ps. Quarter ($\lambda/4$) and half wave plates ($\lambda/2$) placed in front of the polarization beam splitter (PBS1) are used to select the signal and idler photon pairs co-polarized with the pump and to reject the Raman-scattering (RS) cross-polarized with the pump \cite{XYLi04}.

Because the fiber nonlinearity is weak, only about 0.04 photon pair is produced within a typical 4 ps duration pump pulse containing $10^{7}$ photons. Thus, to reliably detect the correlated photon pairs, a high pump-to-signal rejection ratio is required. For HSPS1, we achieve this by passing the output of NLI through a dual-band filter (DBF1), which can separate the photon pairs from the pump with an isolation greater than $\sim 110$ dB. The DBF is realized by cascading a tunable filter and a programmable optical filter (POF, model: Finisar Waveshaper 4000S), and the spectrum of DBF in each band is rectangularly shaped. The bandwidth of pass band in both signal and idler fields can be adjusted from 0.1 nm to 1.5 nm, and the central wavelength of each band can be tuned from 1530 to 1570 nm.  The signal and idler photons passing through DBF1 are detected by single photon detectors, SPD1 and SPD2, which are operated at gated Geiger mode at a rate that is the same as the repetition rate of pump, and the dead time of the gate is set to be 10 $\mu$s.
The detection efficiency of each SPD is $\sim15\%$.  The detection events of a photon in signal (idler) band heralds the presence of a single photon in idler (signal) band with the probability of $h_{i(s)}$.


We first verify the spectral property of photon pairs generated by the three-stage NLI. In the experiment, the pump power of P$_1$ is 50 $\mu$W, the bandwidth of DBF in both signal and idler bands is 0.1 nm. The two outputs of DBF1 (idler1 and signal1 fields) are directly coupled into SPD1 and SPD2, respectively. We measure both the two-fold coincidence counts of signal and idler photons originated from the same pump pulse and adjacent pulses, $C^{c}$ and $C^{acc}$, when the central wavelengths of DBF in signal (idler) channel are scanned from 1552.5 (1545) to 1558 (1539.5) nm with a step of 0.1 nm.
Then, we deduce the true coincidence counts of photon pairs $C^{T}$ by subtracting the measured $C^{acc}$ from $C^{c}$. Fig. 2(b) plots the contour map of $C^{T}$ in the wavelength coordinates of $\lambda_s$ and $\lambda_i$, which reflects the JSI of  photon pairs. One sees that the contour map exhibits “islands” pattern and agrees well with the theoretical prediction in Fig. 1(c).

\begin{figure}[htb]
	\centering
	\includegraphics[width=0.48\textwidth]{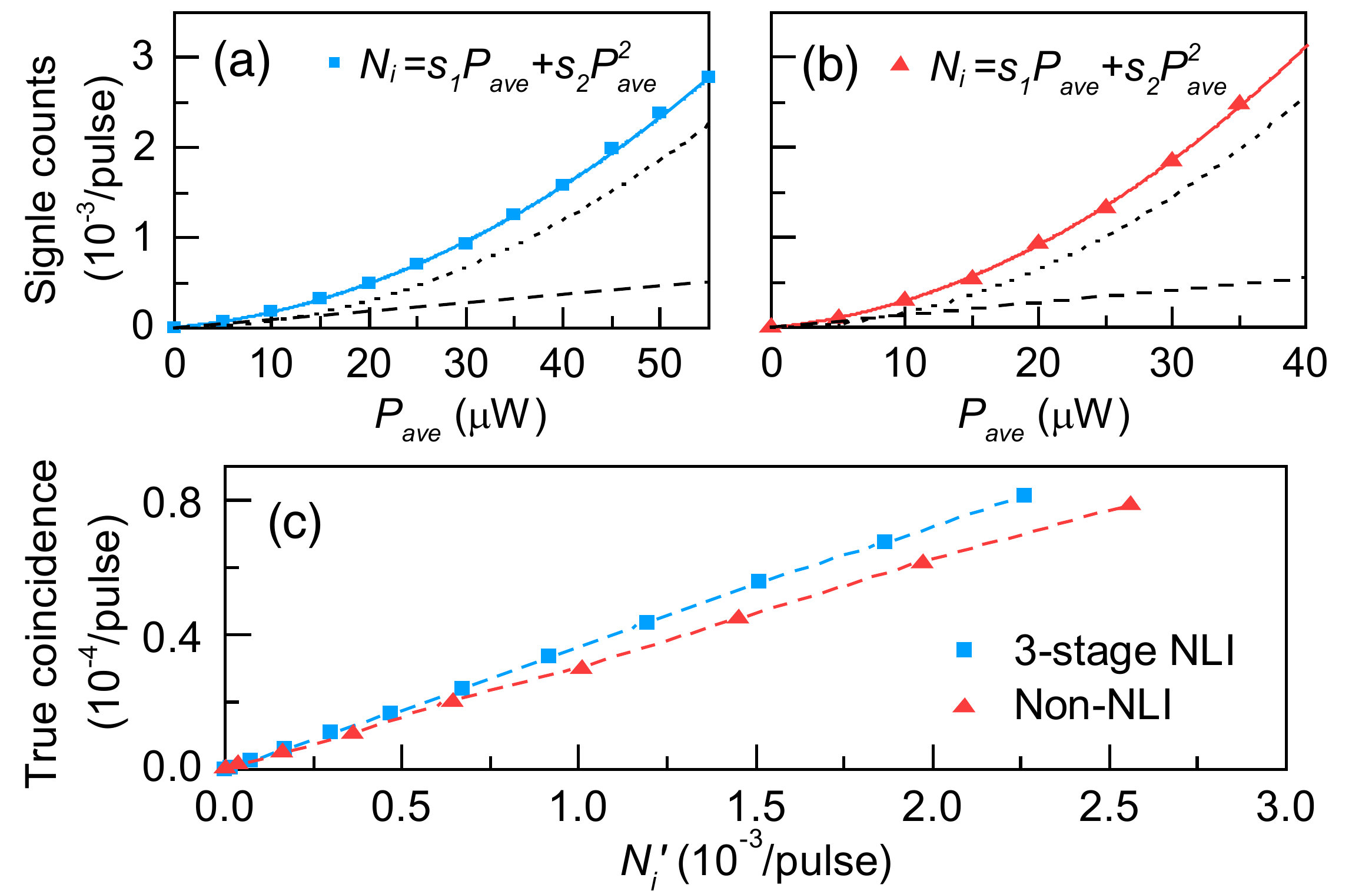}
	\caption{(a) and (b) are the counting rates of SPD1 for individual idler band versus the average pump power $P_{ave}$ for the three-stage NLI and non-NLI (single piece 450-m long DSF) cases, respectively. The second-order polynomials  $N_{i}=s_1 P_{ave} + s_2 P_{ave}^2$ (solid curves) are used to fit the data in plots (a) and (b). The linear and quadratic terms of the fitting functions are represented by the dashed and dotted lines, respectively. (c) The true coincidence rate versus the counting rate $N'_i=s_2 P_{ave}^2$ of idler photons originated from FWM in two cases.}
	\label{fig-single_coin}
\end{figure}

We then measure the achievable heralding efficiency of the HSPS1. Note that in Fig. 2(b), the shape of island centering at 1543.8 and 1553.7 nm is round, indicating the spectral factorable property. By properly adjusting the DBF1, we select out this round shaped island to efficiently obtain uncorrelated photon pairs. In this experiment, the bandwidth of each band is adjusted to be 1.5 nm (see the dashed lines in Fig. 2(b)), and the overall detection efficiencies (including the efficiencies of SPD and DBF1) in signal and idler bands are $\eta_s = 4.1\%$ and $\eta_i = 4.3\%$, respectively.
During the measurement, we record the single counts of individual signal and idler fields and the two-fold coincidence counts of two fields under different pump power. For each pump power level, we subtract $C^{acc}$ to obtain the true coincidence counting rate $C^{T}=C^{c}-C^{acc}$. The measured data of single counts in signal (idler) band is fitted with $N_{s(i)}=s_1 P_{ave} + s_2 P_{ave}^2$, where $P_{ave}$ is the average pump power, $s_1$ and $s_2$ are the fitting parameters. $s_1 P_{ave}$ and $s_2 P_{ave}^2$ respectively correspond to the intensities of spontaneous Raman scattering and FWM in NLI. The results are shown in Fig. \ref{fig-single_coin}(a) where the dashed and  dotted lines are $s_1 P_{ave}$ and $s_2 P_{ave}^2$, respectively.
We then deduce the collection efficiency of signal (idler) photons according to $h_{s(i)}= \frac{C^{T}}{\eta_{i(s)}N'_{i(s)}}$, where $N'_{s(i)}=s_2 P_{ave}^2$ is the quadratic term of $N_{s(i)}$ representing the production rate of photon pairs \cite{LY11, su-arxiv}.
The measured heralding efficiency $h_s$ and $h_i$ are about 91.2$\%$  and $90.5 \%$. As a comparison,
we repeat the efficiency measurement (Fig. \ref{fig-single_coin}(b)) by replacing the NLI with a single piece 450-m long DSF (referred to as non-NLI case), which is realized by removing the SMFs placed in NLI and connecting the three DSFs. From the results of measurement, we find that the collection efficiency of $h_s$ and $h_i$ are about $67.1\%$ and $70.1\%$ for the non-NLI case. Figure 3(c) clearly illustrates that the collection efficiency of signal and idler photon pairs is improved by the NLI.


The detection event of SPD1 in idler output band of NLI is expected to
herald the presence of single photons in the signal band. The photon statistical property of HSPS1, is characterized by $g^{(2)}$, the intensity correlation function of the heralded single photons. To measure $g^{(2)}$, the heralded photons in signal1 field are sent into
a Hanbury Brown-Twiss (HBT) interferometry consisting of a beam splitter (BS2) and two
detectors (SPD2 and SPD3). During the measurement, we record the single counts of SPD1 ($N_i$),
three-fold coincidence counts of SPD1, SPD2 and SPD3 ($N_{123}$), and two-fold coincidence counts between SPD1 and SPD2 (SPD3) ($N_{12}$ and $ N_{13}$). The value $g^{(2)}$ is deduced from the relation: $g^{(2)}=\frac{N_{123} N_{i}}{N_{12} N_{13}}$. Fig. 4(a) shows the measured $g^{(2)}$ as a function of pump power. One sees that $g^{(2)}$ increases with the pump power. This is due to multi-pair events whose probability increases as the pump power or production rate of photon pairs increases.
At a pump power of 30 $\mu$W, corresponding to photon pair production rate of about 0.015 pairs/pulse (after correcting Raman scattering), $g^{(2)}$ of HSPS1 is about $0.034 \pm 0.012$, which is significantly less than the classical limit of $1$. Even when the production rate is increased to 0.043 pairs/pulse (under the pump power of 50 $\mu$W), the value of $g^{(2)}=0.219\pm0.008$ is still well below the classical limit of $1$, which clearly illustrates the non-classical nature of our single photon source.

Finally, we verify the modal purity of HSPS1 by building another HSPS (HSPS2) and performing two-photon HOM interference measurement between the two independent HSPSs (see Fig. 2(a)). The two HSPSs are identical. The detection event of SPD1 (SPD4) in idler1 (idler2) field heralds the presence of a single photon in signal1 (signal2) field. In this experiment, the outputs of the two HSPSs (signal1 and signal2) are carefully path matched and simultaneously fed
into a 50/50 fiber beam splitter (BS2) from two input ports, respectively. Before
sending to BS2, the output of HSPS2 (signal2 field) is delayed by a translation stage. To ensure the two input fields of BS2 have
the identical polarization, the polarization of signal2 field is properly adjusted by a fiber polarization controller (FPC). Note that SPD3 and SPD4 are the same as SPD1 and SPD2, except that their quantum efficiencies are slightly different. During the measurement, the pump powers of both P1 and P2 are about 50 $\mu$W, and the production rate of heralded single photons is about 0.039 photons/pulse. The overall detection efficiencies for the signal and idler photons (including insertion loss of DBPF1 or DBPF2) is $\sim5\%$.
We record the four-fold coincidences of four SPDs versus the position of the translation stage. The raw data (after subtracting the dark counts of SPDs) is plotted in Fig. 5(a), showing the visibility of the HOM interference is 81$\% \pm 6 \%$.

For an ideal single photon state, there is no need to modify the measured four-fold coincidences if the dark counts of the SPDs have been subtracted. But for the heralded single photon state generated from fiber, background counts caused by Raman scattering and the multi-pair events originated from FWM should be subtracted as well. We first correct the influence of Raman scattering, whose photon number can be extracted from the linear term ($s_1 P_{ave}$ ) in Fig. 3(a). Since the strength of the Raman scattering is quite weak, the impact of Raman scattering upon the visibility $V$ is similar to that of the dark counts of SPDs. It is straightforward to obtain the visibility after correction of Raman effect: $V\approx (90 \pm 6)\%$. We then analyze the degradation of $V$ due to multi-pair events by using Eq. (12) in Ref. [\onlinecite{Ma15}] according to the deduced brightness of HSPS (0.039 photons/pulse) in the measurement and find the result after correction is $V \approx (95 \pm 6)\%$.

In addition to the HOM interference measurement, we characterize the mode number of individual signal field for NLI by measuring its intensity correlation function $g_s^{(2)}$. In this measurement, the pump of HSPS2 (P2) is blocked and only the two-fold coincidence counts of SPD2 and SPD3 are analyzed by our counting system. The directly measured result is $g_s^{(2)}=1.8$. But after correcting the contribution from Raman scattering\cite{Su19}, we find the intensity correlation function for the signal field via FWM is $g^{(2)}_s=1.96$ \cite{Su19,Su19_PRA}, from which the mode number of $M=1.04$ can be deduced with  $M=\frac{1}{g^{(2)}_s -1}$. Hence, the corrected visibility $V$ and deduced mode number $M$ are consistent with relation $V=1/M$.

\begin{figure}[htbp]
	\centering
	\includegraphics[width=0.48\textwidth]{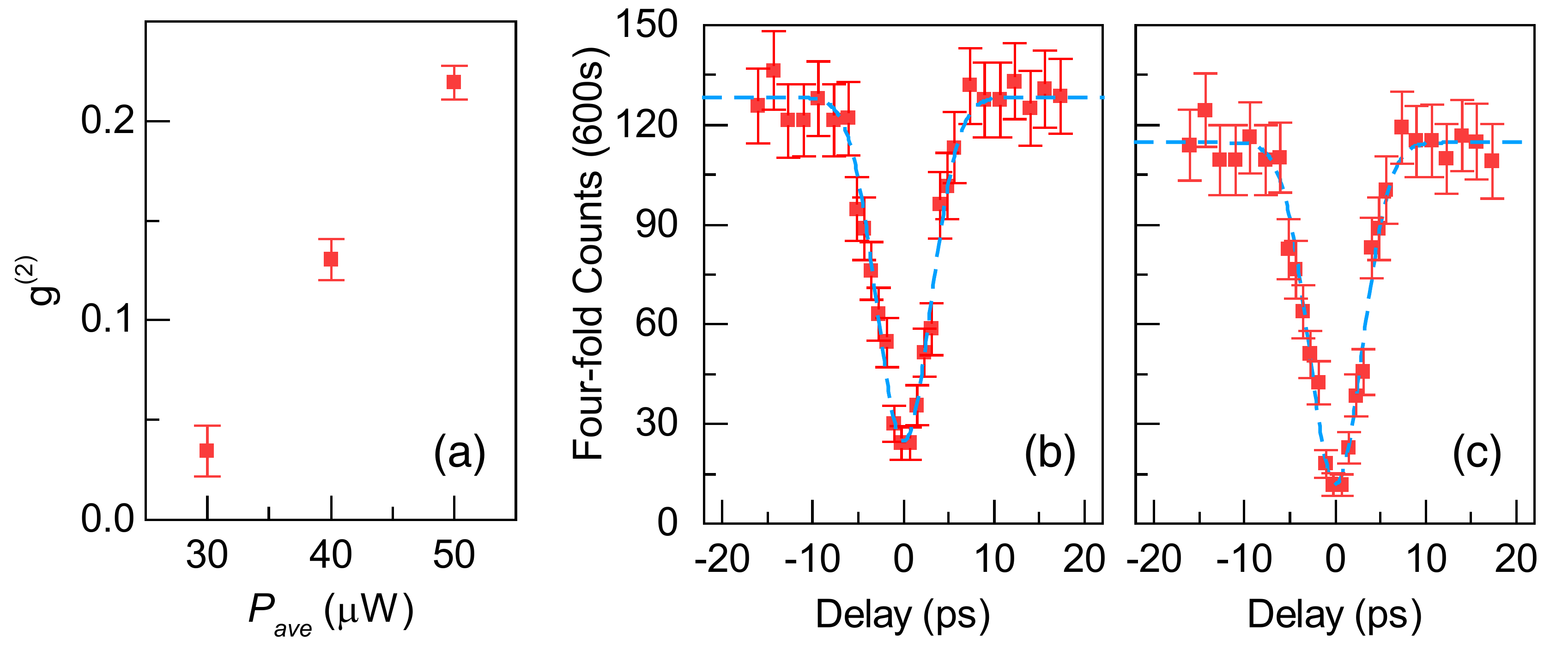}
	\caption{(a) Intensity correlation function $g^{(2)}$ of HSPS versus the average pump power $P_{ave}$. (b) The raw data (with only dark counts of SPDs subtracted) and (c) corrected data (Raman noise subtracted) of four-fold coincidences as a function of the relative delay between the two HSPS. The dashed line is the fitting to a Gaussian function.}
	\label{fig-HOM-result}
\end{figure}

In conclusion, we experimentally study a NLI containing three nonlinear media for the first time. Our investigation indicates that the performance of pulse-pumped three-stage NLI is better than that of a two-stage NLI \cite{Su19} in engineering the spectral property of photon pairs. We demonstrate that the three-stage NLI  can be used to develop pure state of heralded single photons with high heralding efficiency. When the brightness and heralding efficiency of each telecom band HSPS are about 0.039 photons/pulse and $90\%$, respectively, the raw data for the visibility of two-photon HOM interference between two independent HSPSs is measured to be  $81\% \pm 6 \%$. After correcting the Raman scattering and  multi-pair events, the visibility can reach  $95\% \pm 6 \%$. Our results show the heralded single photons generated from the three-stage NLI simultaneously have the advantages of high purity, high heralding efficiency, high brightness, and flexibility in wavelength and bandwidth selection. Moreover, we note that the performance of our HSPSs is robust to the structure fluctuations of NLI, because we do not observe any change in the key performance parameters of the HSPSs when the lengths of DSF and SMF in NLI are varied within a few centimeters.
We believe that the fiber-based multi-stage nonlinear interferometer, which is easy to implement, will be a useful resource for quantum information processing involving quantum interference among multiple independent sources.

The work is supported in part by the National Natural Science Foundation of China (11527808, 91736105, 11874279).


%

\end{document}